# ON BOSONIC SOLUTIONS OF THE FOLDY – WOUTHUYSEN AND DIRAC EQUATIONS FOR FREE FIELD

[1]Simulik V.M., [1]Krivsky I.Yu., [2]Zajac T.M.

[1]Institute of Electron Physics, National Academy of Sciences of Ukraine. 21 Universitetska Str., 88000 Uzhgorod, Ukraine
e-mail: vsimulik@gmail.com
[2]Department of Electronic Systems, Uzhgorod National University, 13 Kapitulna Str., 88000 Uzhgorod, Ukraine

**Abstract.** The consideration of the Fermi – Bose duality of the Dirac equation with arbitrary mass has been continued. The bosonic solutions of this equation in the Foldy – Wouthuysen representation are found. The Bargman – Wigner analysis of additional solutions and comparison with standard fermionic solutions are carried out. The evident demonstration of additional bosonic features of the Foldy – Wouthuysen and Dirac equation is presented.

## 1. Introduction

Starting from the first steps of quantum mechanics and during the period of its growth many authors are being investigating the Fermi – Bose duality of the massless spinor field and of the related electromagnetic field in the terms of field strengths (see the relevant references in [1]). Nowerdays the interest to this problem is still expressed in different publications (see e. g., the articles [1–5] and the references therein).

The Fermi – Bose duality of the spinor field of arbitrary mass was considered in [6–9] but only for the case of stationary Dirac equation. Only recently the bosonic characteristics of the general, time dependent (non-stationary), Dirac equation with nonzero mass were found in papers [1, 10] (see also e-preprint [11]).

This paper is the direct continuation of the article [1]. Recall briefly that in [1], with the purpose of finding the bosonic symmetries of the Dirac equation with $m \neq 0$, the 64-dimensional extended real Clifford – Dirac algebra (ERCD algebra in the notations of [1]) was put into consideration. Further, in [1], as subalgebra of the ERCD algebra, the wide 32-dimensional $A_{32}$ algebra of invariance of the Dirac equation with arbitrary mass was found. On the basis of the part of $A_{32}$ algebra, the hidden invariance of this equation with respect to bosonic representation of the main relativistic group – of universal covering $\mathcal{P} \supset \mathcal{L} = \mathrm{SL}(2,\mathrm{C})$ of proper ortochronous Poincare group $\mathrm{P}_+^\uparrow \supset \mathrm{L}_+^\uparrow = \mathrm{SO}(1,3)$ – was found. Thus, in our publications [1, 11], the bosonic (spin $s=(1,0)$) symmetries of the Dirac equation with nonzero mass were put into consideration.

Following the relationship between the symmetries and the corresponding solutions, we are able now to add to the result of [1] the explicit form of bosonic solutions of the Dirac equation for the arbitrary mass. We prove below that the set of such solutions is invariant with respect to the bosonic $\mathcal{P}^{\mathrm{B}}$ representation of the group $\mathcal{P}$, which is generated by the bosonic spin s=(1,0) representation of the SU(2) group. By comparing the bosonic solutions with the standard spin $s = \frac{1}{2}$-doublet solutions of the Dirac equation and according to the Bargman-Wigner analysis of both solutions, the evident demonstration of the additional bosonic characteristics of the Dirac equation is given.

Note that for generalization of the results of [1] it is convenient (similarly to [1]) to start from the canonical Foldy – Wouthuysen (FW) representation [12] of the spinor field. The corresponding results in the standard local representation of the Dirac equation may be found very easily with the help of well-known FW transformation [12]. The reasons of such start are the following. It is the canonical FW representation, in which the mathematical and physical aspects of the spinor field theory are most evident and adequate (see, e. g. [12]). Therefore, the additional mathematical formalism – the ERCD algebra – was put into consideration in [1, 11], namely in the FW representation but not in the standard local Pauli – Dirac (PD) representation. Let us recall that ERCD algebra includes the ordinary 16-dimensional Clifford – Dirac algebra as a subalgebra, $ERCD \supset CD$.

Below for our purposes we also use essentially the start from the canonical FW representation [12] and the formalism of the ERCD algebra still is put into consideration namely in this representation.

## 2. Notations and definitions

We use the system of units $\hbar = c = 1$, the metric $g = (g^{\mu\nu}) = (+ - - -)$, $a^\mu = g^{\mu\nu} a_\nu$, the Greek indices are changed in the region of 0,1,2,3, Latin – $\overline{1,3}$, and summation over twice repeated index is implied.

The mathematically well-defined consideration and adequate physical formulation demand, for the solutions of the FW equation, to determine precisely not only the domain of definition and the range of values, but also the manifold (class), which these solutions belong to. This determination demands to consider all solutions (both Fermi and Bose) in one and the same quantummechanical 4-component Hilbert space

$$H^{3,4} = L_2(R^3) \times C^4, \quad L_2(R^3) = \{\phi: R^3 \to C^4; \int d^3x |\phi(x)|^2 < \infty\}, \quad (1)$$

which is included into the rigged Hilbert space

$$S^{3,4} \subset H^{3,4} \subset {}^\times S^{3,4}. \quad (2)$$

Here the symbol «×» in ${}^\times S^{3,4}$ means that the space of generalized functions ${}^\times S^{3,4}$ is conjugated by the corresponding topology to the Schwartz test functions space $S^{3,4}$.

Note that the Schwartz test functions space $S^{3,4}$ in (2) is kernel, i. e it is dense both in the space $H^{3,4}$ and in the space ${}^\times S^{3,4}$ of generalized Schwartz functions. Note also that the space $S^{3,4}$ (as well as the space $H^{3,4} \subset {}^\times S^{3,4}$) is invariant with respect to the Fourier transformation; therefore, the coordinate ($\vec{x}$-) and momentum $\vec{k}$-realization of these spaces are alternative. Further, the choice of the space $S^{3,4}$ as the domain of definition of the operators of physical values is caused by the fact that simultaneously it is also the range of their values. All these facts enable one not to appeal to the mathematically complicated functional formalism of the space ${}^\times S^{3,4}$ and to fulfill all necessary calculations in the subspace $S^{3,4} \subset H^{3,4}$ without any loss of generality and mathematical correctness. Some more details of our motivation in such choice see e. g. in [13].

As we are going to explicit consideration and direct comparison of the different types of solutions, let us recall at first the well-known fermionic solutions of the Dirac equation in FW representation. Simultaneously we shall put into consideration and explore the conceptions and notations, which will be useful for us and are necessary for the construction of additional solutions and symmetries. For example, similarly to [1], we shall appeal to the prime anti-

Hermitian generators $q = -i\hat{q}$ of corresponding groups and their representations. The relationship with standard Hermitian $\hat{q}$ generators is useful to be demonstrated on the example of the well-known momentum and spin operators

$$p_\mu = \partial_\mu = -i\hat{p}_\mu, \quad s^{\mu\nu} = -i\hat{s}^{\mu\nu} = \frac{1}{4}[\gamma^\mu, \gamma^\nu], \tag{3}$$

where ordinary Hermitian momentum and spin generators are given by

$$\hat{p}_\mu = i\partial_\mu, \quad \hat{s}^{\mu\nu} = \frac{i}{4}[\gamma^\mu, \gamma^\nu]. \tag{4}$$

Here for definiteness the explicit form of the $\gamma^\mu$ matrices is taken in standard PD representation, see e. g. formulae (5) in [1]. Indeed, these *prime* anti-Hermitian generators (3) have a direct physical meaning and are related to the real parameters $a = (a_\mu)$, $\varpi = (\varpi_{\mu\nu})$ of 4-translations and 4-rotations. Note that using the prime anti-Hermitian generators in the group theory does not contradict the general formalism, see e. g. special remarks in [14].

The FW equation

$$(\partial_0 + i\gamma^0\omega)\phi(x) = 0; \quad \omega \equiv \sqrt{-\Delta + m^2}, \quad x \in M(1,3), \quad \phi \in H^{3,4}; \tag{5}$$

has the wide symmetry properties. In [1, 11], the pure matrix 32-dimensional $A_{32} \subset$ ERCD algebra of invariance of this equation was found. The set of corresponding conservation laws for the spinor field is one of the consequences of this algebra. Note that the final explicit form of any conservation law for the spinor field does not depend on the representation (canonical FW or standard local PD), in which the Dirac equation is considered. Below we analyze only such symmetries and corresponding additional solutions of the equation (5) (and briefly – the corresponding conservation laws), which are directly related to physically meaningful Fermi – Bose duality of the spinor field.

In order to demonstrate the Fermi – Bose duality of the Dirac equation in FW representation (5) it is useful to pay attention to the subalgebra SO(6) of the complete algebra of invariance $A_{32}$ of equation (5). Representation of SO(6) in the space $H^{3,4}$ (which includes, as a subalgebra, the algebra of the group of rotations SO(3) in the space $R^3 \subset M(1,3)$) is generated according to formulae (10) in [1] by the pure matrix operators ($\gamma^A$ matrices, $A = \overline{1,6}$):

$$\alpha^{A6} = \gamma^A: \gamma^1, \gamma^2, \gamma^3, \gamma^4 \equiv \gamma^0\gamma^1\gamma^2\gamma^3, \gamma^5 \equiv \gamma^1\gamma^3 C, \gamma^6 \equiv i\gamma^1\gamma^3 C; \gamma^A\gamma^B + \gamma^B\gamma^A = -2\delta^{AB}. \tag{6}$$

Here is the operator of complex conjugation, $C\phi = \phi^*$ (i. e. is the involution operator in the space $H^{3,4}$), and, further, is the ort of ERCD algebra [1].

**Remark 1.** Note that the 64-dimensional ERCD algebra was put into consideration in [1, 11] as a complete set of operators, which are the products of operators $i = \sqrt{-1}$, and 16 orts of the standard CD algebra. In the ERCD algebra, 7 orts have the properties of 5 generators $\gamma^0, \gamma^1, \gamma^2, \gamma^3, \gamma^4 \equiv \gamma^0\gamma^1\gamma^2\gamma^3$ of standard CD algebra. It is convenient to chose 7 orts of the ERCD-algebra, which are satisfied the Clifford commutation relations $\gamma^A\gamma^B + \gamma^B\gamma^A = -2\delta^{AB}$, AB=$\overline{1,7}$, in the form of the set of 6 matrices (6) together with the matrix $\gamma^7 = i\gamma^0$. Therefore, the standard 16-dimensional CD algebra (SO(1,5) algebra) is generalized to the 28-dimensional $CD_{ER}$ algebra (with N=28 independent orts), which is isomorphic to SO(8) or SO(1.7) algebra, see [1]. Corresponding generators

$$s^{\tilde{A}\tilde{B}} = \{s^{AB} \equiv \frac{1}{4}[\gamma^A, \gamma^B], s^{A8} = -s^{8A} = \frac{1}{2}\gamma^A\}, \quad \tilde{A},\tilde{B}=\overline{1,8}, \tag{7}$$

satisfy the commutation relations

$$[s^{\tilde{A}\tilde{B}}, s^{\tilde{C}\tilde{D}}] = \delta^{\tilde{A}\tilde{C}}s^{\tilde{B}\tilde{D}} + \delta^{\tilde{C}\tilde{B}}s^{\tilde{D}\tilde{A}} + \delta^{\tilde{B}\tilde{D}}s^{\tilde{A}\tilde{C}} + \delta^{\tilde{D}\tilde{A}}s^{\tilde{C}\tilde{B}} \tag{8}$$

(in the case of the SO(8) algebra). Namely due to the more extended possibilities of the 28-dimensional $CD_{ER}$ algebra SO(8) (or SO(1.7)), in paper [1] and below the additional bosonic characteristics of the massive spinor field can be found. □

Let us remind that the standard CD algebra with N=16 independent orts is isomorphic to the algebras SO(3.3) [15, 16], SO(1.5) and SO(6) [1]. For our constructions it is useful to appeal to the SO(1.5) and SO(6) forms of the CD algebra, see., e. g., [1].

## 3. Fermionic solutions of the Foldy – Wouthuysen equation

In the next section we shall consider the Bose solutions of the FW equations. To be able to compare such additional solutions with the well-known Fermi solutions, here we briefly consider the standard solutions. Following this procedure, we determine the formalism and the method, which are necessary to us. Thus, below the explicit form of the Fermi solution of the Dirac equation (both in the FW and PD representations) is given. For the same purposes in Section 5 the Bargman-Wigner analysis of fermionic features of this solution is presented.

The Fermi solution of equation (5) is determined with the help of some stationary complete set of operators of fermionic physical quantities for the spin $s = \frac{1}{2}$-doublet ($\frac{1}{2}$D) in the FW representation, i. e., the set "momentum, sign of the charge, projection $s^3$ of the spin $\vec{s}$":

$$(\vec{p} = -\nabla,\ i\gamma^0,\ s^3), \tag{9}$$

where

$$s^3 \in \vec{s} = (s^1 = s^{23},\ s^2 = s^{31},\ s^3 = s^{12}) = \frac{1}{2}(\gamma^2\gamma^3,\ \gamma^3\gamma^1,\ \gamma^1\gamma^2) \Rightarrow \vec{s}^2 = -\frac{1}{2}\left(\frac{1}{2}+1\right)I_4, \tag{10}$$

and $I_4$ is unit $4\times 4$-matrix.

The fundamental solutions of equation (5), which are the common eigenstates of the fermionic complete set (9), have the form

$$\varphi^{\varepsilon_1}_{\vec{k}\varepsilon_2}(t,\vec{x}) = \frac{1}{(2\pi)^{3/2}} e^{i\varepsilon_1(\tilde{\omega}t - \vec{k}\vec{x})} D^{\varepsilon_1}_{\varepsilon_2}, \tag{11}$$

where $\varepsilon_1$ is the eigenvalue of the sign of the charge operator for the spinor field, $\frac{\varepsilon_2}{2}$ are the eigenvalues $\left(\frac{1}{2}, -\frac{1}{2}, \frac{1}{2}, -\frac{1}{2}\right)$ of the spin projection $s^3$ operator of the spin $\vec{s}$ (10), and $D^{\varepsilon_1}_{\varepsilon_2}$ are the Cartesian orts in the space $H^{3,4}$ (1):

$$D^{\varepsilon_1}_{\varepsilon_2}:\ D^-_+ = d_1,\ D^-_- = d_2,\ D^+_+ = d_3,\ D^+_- = d_4;\ d_\alpha = (\delta^\beta_\alpha). \tag{12}$$

Solutions (11) are the generalized states belonging to the space $^\times S^{3,4}$. Nevertheless, they form the complete orthonormalized system of generalized states. Therefore, any fermionic physical state of the FW field $\phi$ from the dense in $H^{3,4}$ manifold $S^{3,4}$ (the general fermionic solution of the equation (5)) is uniquely presented in the form

$$\phi_{\frac{1}{2}D}(x) = \frac{1}{(2\pi)^{3/2}} \int d^3k[e^{-ikx}(a^-_+(\vec{k})D^-_+ + a^-_-(\vec{k})D^-_-) + e^{ikx}(a^{*+}_+(\vec{k})D^+_+ + a^{*+}_-(\vec{k})D^+_-)];\ kx = \tilde{\omega}t - \vec{k}\vec{x}, \tag{13}$$

where, for $\frac{1}{2}$D, $a^\alpha_\beta(\vec{k}) \equiv a^{\varepsilon_1}_{\varepsilon_2}(\vec{k})$ are the momentum-charge-spin quantummechanical amplitudes of the probability distribution according to the eigenvalues of the stationary complete set (9) (note, that the concept of degeneration for the arbitrary stationary complete set of operators is absent). We pay attention that for $\forall \phi(x)$ (13), belonging to the space $S^{3,4} \subset H^{3,4}$, the amplitudes $a^{\varepsilon_1}_{\varepsilon_2}(\vec{k})$ belong to the Schwartz complexvalued test function space $S(R_{\vec{k}}) \subset L_2(R_{\vec{k}})$, where $R_{\vec{k}}$ is the spectrum of the operator $\vec{p}$ from (9).

The fermionic general solution of the type (13), which is related to any other fermionic stationary complete set of operators, e.g., to momentum-charge-helicity complete set

$$\left(\vec{p} = -\nabla, \; i\gamma^0, \; h = \frac{\vec{s} \cdot \nabla}{|\nabla|}\right), \qquad (14)$$

is constructed similarly.

In the PD representation, solution (13) of the standard Dirac equation

$$(\partial_0 + i\hat{H})\psi(x) = 0; \quad \hat{H} \equiv \gamma^0(\vec{\gamma} \cdot \hat{\vec{p}} + m), \; \hat{\vec{p}} = -i\nabla, \qquad (15)$$

is given by the well-known expression

$$\psi(x) = V^{-1}\phi_{\frac{1}{2}D}(x) = \frac{1}{(2\pi)^{3/2}} \int d^3k [e^{-ikx}(a_+^-(\vec{k})\mathrm{v}_+^- + a_-^-(\vec{k})\mathrm{v}_-^-) + e^{ikx}(a_+^{*+}(\vec{k})\mathrm{v}_+^+ + a_-^{*+}(\vec{k})\mathrm{v}_-^+)]. \qquad (16)$$

Here the 4-component spinors $\mathrm{v}_{+-}^{-+}$ have the form

$$\mathrm{v}_{\varepsilon_2}^- = V^{-1}\mathrm{D}_{\varepsilon_2}^- = N \begin{vmatrix} (\tilde{\omega}+m)\mathrm{d}_{\varepsilon_2} \\ (\vec{\sigma} \cdot \vec{k})\mathrm{d}_{\varepsilon_2} \end{vmatrix}, \; \mathrm{v}_{\varepsilon_2}^+ = V^{-1}\mathrm{D}_{\varepsilon_2}^+ = N \begin{vmatrix} (\vec{\sigma} \cdot \vec{k})\mathrm{d}_{\varepsilon_2} \\ (\tilde{\omega}+m)\mathrm{d}_{\varepsilon_2} \end{vmatrix}, \qquad (17)$$

where $N^{-1} = \sqrt{2\tilde{\omega}(\tilde{\omega}+m)}$, nd $\mathrm{d}_{\varepsilon_2}$ are the 2-component eigenstates of the operator $\frac{1}{2}\sigma^3$ (the explicit form of the FW operator $V^{-1}$ see, e.g., in the formula (3) of [1]).

Note that in the PD representation the 4-component spinors $\mathrm{v}_{+-}^{-+}$ (17) are not the eigenvectors of the spin projection operator $s^3$ (10). They are the eigenvectors of the nonlocal $s^{3\mathrm{Dirac}}$ correct spin projection operator $\vec{s}^{\mathrm{Dirac}} = V^{-1}\vec{s}V$ only; the explicit form of the $\vec{s}^{\mathrm{Dirac}}$ operator see, e.g., in [12, 13].

## 4. Bosonic solutions of the Foldy – Wouthuysen equation

In order to construct the bosonic stationary complete set of operators, which commute with the operator $(\partial_0 + i\gamma^0\omega)$ of equation (5), we use the part of the generators of the SO(6) algebra (see formulae (10) in [1]), i.e., the two sets of the SU(2) generators. The first set of the SU(2) generators is known well. They are the standard fermionic spin operators (8). Another set of the SU(2) generators is given by the formulae

$$\vec{s}' = (s'^1 = s'^{23}, \; s'^2 = s'^{31}, \; s'^3 = s'^{12}) = \frac{1}{2}(-\gamma^0\gamma^2 C, \; i\gamma^0\gamma^2 C, \; -i) \Rightarrow \vec{s}'^2 = -\frac{1}{2}\left(\frac{1}{2}+1\right)\mathrm{I}_4, \qquad (18)$$

(here, as well as above in (6), is the complex conjugation operator in $\mathrm{H}^{3,4}$). The sets of the SU(2) generators $\{\vec{s}\}$ (10) and $\{\vec{s}'\}$ (18) commute between each other, $[\{\vec{s}\}, \{\vec{s}'\}] = 0$, and are linked by the operator $u$,

$$\vec{s}' = u\vec{s}u^{-1}, \; u = \begin{vmatrix} 0 & 0 & 0 & C \\ 1 & 0 & 0 & 0 \\ 0 & C & 0 & 0 \\ 0 & 0 & 1 & 0 \end{vmatrix}, \; u^{-1} = \begin{vmatrix} 0 & 1 & 0 & 0 \\ 0 & 0 & C & 0 \\ 0 & 0 & 0 & 1 \\ C & 0 & 0 & 0 \end{vmatrix}, \; uu^{-1} = u^{-1}u = 1. \qquad (19)$$

Note that operator $u$ (19) does not change the explicit form of the Hamiltonian $i\gamma^0\omega$ of equation (5), i.e., $ui\gamma^0\omega u^{-1} = i\gamma^0\omega$. Therefore, the explicit form of the FW equation (5) does not change under the transformation (19)

The spin SU(2) generators $\underset{\sim}{\vec{s}}$ for the bosonic stationary complete set in the Cartesian basis are found in the following way

$$\underset{\sim}{\vec{s}} = W(\vec{s} + \vec{s}')W^{-1}, \qquad (20)$$

where the operator $W$ of corresponding transformation has the form

$$W = \frac{1}{\sqrt{2}} \begin{vmatrix} \sqrt{2} & 0 & 0 & 0 \\ 0 & -1 & 0 & C \\ 0 & 0 & -i\sqrt{2} & 0 \\ 0 & -C & 0 & -1 \end{vmatrix}, \quad W^{-1} = \frac{1}{\sqrt{2}} \begin{vmatrix} \sqrt{2} & 0 & 0 & 0 \\ 0 & -1 & 0 & -C \\ 0 & 0 & i\sqrt{2} & 0 \\ 0 & C & 0 & -1 \end{vmatrix}, \quad WW^{-1} = W^{-1}W = 1. \quad (21)$$

Note that without transformation $W$ (20), (21) the operators $(\vec{s} + \vec{s}')$ also satisfy the commutation relations for the prime generators of the SU(2) algebra, but the Casimir operator $(\vec{s} + \vec{s}')^2$ is not diagonal. The transformation $W$ (20), (21) does not change the explicit form of the spin projection $s^3 + s'^3$ and SU(2) commutation relations, but provide the diagonalization of the operator $(\vec{s} + \vec{s}')^2$. Thus, the generators $\underset{\sim}{\vec{s}}$ (18) are found as

$$\underset{\sim}{s}^1 = \frac{1}{\sqrt{2}}\begin{vmatrix} 0 & i & 0 & 0 \\ i & 0 & C & 0 \\ 0 & -C & 0 & 0 \\ 0 & 0 & 0 & 0 \end{vmatrix}, \underset{\sim}{s}^2 = \frac{1}{\sqrt{2}}\begin{vmatrix} 0 & 1 & 0 & 0 \\ -1 & 0 & -iC & 0 \\ 0 & iC & 0 & 0 \\ 0 & 0 & 0 & 0 \end{vmatrix}, \underset{\sim}{s}^3 = \begin{vmatrix} -i & 0 & 0 & 0 \\ 0 & 0 & 0 & 0 \\ 0 & 0 & -i & 0 \\ 0 & 0 & 0 & 0 \end{vmatrix}, \underset{\sim}{\vec{s}}^2 = -1(1+1)\begin{vmatrix} I_3 & 0 \\ 0 & 0 \end{vmatrix}, \quad (22)$$

where $I_3$ is the unit $3 \times 3$ matrix.

It is interesting to note that the operator $W$ (20), (21) (as well as the operator $u$ (19)) does not change $i\gamma^0 \omega$, i.e., $Wi\gamma^0\omega W^{-1} = i\gamma^0\omega$. Therefore, the FW equation (5) does not change its explicit form (5) under the transformation from the fermionic representation to the Bosonic one.

Hence, a bosonic solution of the FW equation (5) is found completely similarly to the procedure of construction of standard fermionic solution, which is presented in the previous section. Thus, the bosonic solution is determined by some stationary complete set of operators of bosonic physical quantities for the spin s=(1,0)-multiplet in the FW representation, e. g., by the set "momentum-spin projection $\underset{\sim}{s}^3$":

$$(\vec{p} = -\nabla, \underset{\sim}{s}^3), \quad (23)$$

where $\underset{\sim}{s}^3$ is given in (22). In comparison with the fermionic complete set (9), the charge sign operator is absent in the set (23). It is due to the fact that the charge of s=(1,0)-multiplet is equal to zero.

The fundamental solutions of equation (5), which are the common eigensolutions of the bosonic complete set (23), have the form

$$\varphi_{\vec{k}\varepsilon}(t,\vec{x}) = \frac{1}{(2\pi)^{3/2}} e^{i(\tilde{\omega}t - \vec{k}\vec{x})} D_\varepsilon, \quad (24)$$

where $\varepsilon$ are the eigenvalues 1,0,-1,0 of the operator $\underset{\sim}{s}^3$ of the spin projection of the spin $\underset{\sim}{\vec{s}}$ (22), and $D_\varepsilon$ are the Cartesian orts in the same space $H^{3,4}$ (1):

$$D_\varepsilon : D_+ = d_1, D_0 = d_2, D_- = d_3, D_{\underline{0}} = d_4; \quad d_\alpha = (\delta_\alpha^\beta). \quad (25)$$

The bosonic solutions (24) (as well as (11)) are the generalized states, belonging to the space $^\times S^{3,4}$; they form the complete orthonormalized system of bosonic generalized states. Therefore, any bosonic physical state of the FW field $\phi$ from the dense in $H^{3,4}$ manifold $S^{3,4}$ (the general bosonic solution of the equation (5)) is uniquely presented in the form

$$\phi_{(1,0)}(x) = \frac{1}{(2\pi)^{3/2}} \int d^3k [e^{-ikx}(\xi_+(\vec{k})D_+ + \xi_0(\vec{k})D_0) + e^{ikx}(\xi_-^*(\vec{k})D_- + \xi_{\underline{0}}^*(\vec{k})D_{\underline{0}})], \quad (26)$$

where $\xi_\varepsilon(\vec{k})$ are the bosonic momentum-spin quantummechanical amplitudes of the probability distribution according to the eigenvalues of the stationary complete set (21) for the bosonic s=(1,0)-multiplet. And if $\phi_{(1,0)}(x) \in S^{3,4}$, then the bosonic amplitudes $\xi_\varepsilon(\vec{k})$ belong to the Schwartz complexvalued test function space $S(R_{\vec{k}}) \subset L_2(R_{\vec{k}})$.

It is useful to present the Bose general solution (26) not in the orts of Cartesian $\{D_\alpha\}$ basis, but in the orts of a cyclic basis

$$\{C_\alpha\}: \quad C_1 = \frac{1}{\sqrt{2}}\begin{vmatrix}1\\i\\0\\0\end{vmatrix}, \quad C_2 = \begin{vmatrix}0\\0\\1\\0\end{vmatrix}, \quad C_3 = \frac{1}{\sqrt{2}}\begin{vmatrix}i\\1\\0\\0\end{vmatrix}, \quad C_4 = \begin{vmatrix}0\\0\\0\\1\end{vmatrix}, \tag{27}$$

where the spin operators of s=(1,0)-multiplet are given by

$$\underline{s}^1 = \begin{vmatrix}0&0&0&0\\0&0&-C&0\\0&C&0&0\\0&0&0&0\end{vmatrix}, \quad \underline{s}^2 = \begin{vmatrix}0&0&C&0\\0&0&0&0\\-C&0&0&0\\0&0&0&0\end{vmatrix}, \quad \underline{s}^3 = \begin{vmatrix}0&-1&0&0\\1&0&0&0\\0&0&0&0\\0&0&0&0\end{vmatrix}, \quad \vec{\underline{s}}^2 = -1(1+1)\begin{vmatrix}I_3&0\\0&0\end{vmatrix}. \tag{28}$$

The consideration of the bosonic states in the cyclic spin basis is useful to provide the investigation of the relationship between the s=(1,0)-multiplet FW field theory and the Maxwell equations in the terms of field strengths $\vec{E}, \vec{H}$ (the details are not the subject of this brief consideration).

The bosonic solution of the equation (5) in the cyclic spin basis (27) have the form

$$\phi_{(1,0)}(x) = \frac{1}{(2\pi)^{3/2}} \int d^3k [e^{-ikx}(b_+(\vec{k})C_1 + b_-(\vec{k})C_3) + e^{ikx}(b_0(\vec{k})C_2 + b_{\underline{0}}(\vec{k})C_4)], \tag{29}$$

and the relationship between the spin orts $\{D_\alpha\}$ and $\{C_\alpha\}$ is given by the operator

$$\underline{s} = U \vec{\underline{s}} U^{-1}, \quad U d_\alpha = C_\alpha, \quad UU^{-1} = U^{-1}U = 1,$$

$$U = \frac{1}{\sqrt{2}}\begin{vmatrix}1&0&iC&0\\i&0&C&0\\0&\sqrt{2}C&0&0\\0&0&0&\sqrt{2}\end{vmatrix}, \quad U^{-1} = \frac{1}{\sqrt{2}}\begin{vmatrix}1&-i&0&0\\0&0&\sqrt{2}C&0\\iC&C&0&0\\0&0&0&\sqrt{2}\end{vmatrix}. \tag{30}$$

Of course, the main features of solution (29) are similar to the features of the bosonic solution (26). Solution (29) is related to the stationary complete set $(\vec{p} = -\nabla, \underline{s}^3)$, in which $\underline{s}^3$ is given in (28). It is important that transformation $U$ (30) (similarly to other transformations (19), (21) considered here) does not change the operator $i\gamma^0 \omega$. Hence, the operator of the FW equation (5) still has the same explicit form $(\partial_0 + i\gamma^0 \omega)$.

Similarly to the situation with fermionic solutions (Section 3), the bosonic solutions of the (26), (29) type also may be constructed for any other bosonic stationary complete set. For example, the momentum-helicity complete set

$$\left(\vec{p} = -\nabla, \; h = \frac{\vec{\underline{s}} \cdot \nabla}{|\nabla|}\right) \tag{31}$$

is useful.

Let us consider briefly the PD representation. The transformation $\phi_{(1,0)} \to \psi_{(1,0)}$ of the bosonic solutions (26), (29) of the FW equation (5) into the local representation (standard PD representation of the Dirac equation (15)) is provided by the same FW operator $V^{-1}$ [12], which transforms the fermionic solutions (see Section 3, formulae (15) – (17)). Hence, $\psi_{(1,0)} = V^{-1}\phi_{(1,0)}$, $V^{-1}(i\gamma^0 \omega)V = \gamma^0(\vec{\gamma} \cdot \vec{p} + m)$, and the details are not the subject of this brief consideration. It is the direct consequence of the fact that all transformations $u$, $W$, $U$ used here in constructing the bosonic solutions do not change the operator $(\partial_0 + i\gamma^0 \omega)$ of the FW equation (5).

## 5. Bargman – Wigner analysis of the solutions under consideration

In order to answer the question "What is the kind of solution, Fermi or Bose?" the following analysis must be provided. (i). The corresponding (fermionic or bosonic) stationary complete set of operators is the first step of indication (corresponding analysis is given in Sections 3,4). (ii). Bargman – Wigner analysis of the representation of the Poincare group $\mathscr{P}$, according to which the corresponding (one or another) set of solutions is invariant, is the second characteristic. Fulfilling such analysis is the purpose of this section. (iii). The final necessary step is obtaining the corresponding (Fermi or Bose) Poincare list of conservation laws (it will be, but very briefly, considered in next section).

Below we use the following mathematical definition of the concept of symmetry (invariance) of the equations of mathematical physics (see, e. g., [17]). Operator $\hat{q}$ in $H^{3,4}$ is called the invariance transformation operator of equation (5), if the set $\Phi = \{\phi\} \subset H^{3,4}$ of solutions of this equation is invariant with respect to the transformation $\hat{q}$, i. e., if $\hat{q}\Phi = \Phi \subset H^{3,4}$. Thus, relativistic invariance of equation (5) is accepted as the invariance of the set of solutions of this equation with respect to the operators of some representation of universal covering $\mathscr{P} \supset \mathscr{L} = \mathrm{SL}(2,\mathrm{C})$ of proper ortochronous Poincare group $\mathrm{P}_+^\uparrow \supset \mathrm{L}_+^\uparrow = \mathrm{SO}(1,3)$. And Bargman – Wigner analysis of the explicit form of Casimir operators of corresponding representations of this group demonstrates what kind (e. g, Fermi or Bose) is the set of solutions under consideration.

Let us recall that the set $\Phi^{\mathrm{F}} \equiv \{\phi_{\frac{1}{2}\mathrm{D}}\}$ of solutions (13) of equation (5) is invariant with respect to $\mathscr{P}^{\mathrm{F}}$-generators D-64 – D-67 of [18], which in the terms of prime generators have the form

$$p_0 = -i\gamma_0\omega,\ p_n = \partial_n,\ j_{ln}^{\mathrm{F}} = x_l\partial_n - x_n\partial_l + s_{ln},$$
$$j_{0k}^{\mathrm{F}} = x_0\partial_k + i\gamma_0\left\{x_k\omega + \frac{\partial_k}{2\omega} + \frac{(\vec{s}\times\vec{\partial})_k}{\omega + m}\right\}, \tag{32}$$

and satisfy the corresponding commutation relations for the $\mathscr{P}$ generators in manifestly covariant form. Operators $(p_\mu, j_{\mu\nu}^{\mathrm{F}})$ (32) with the help of well known exponential series, which is converged in the space $S^{3,4} \subset H^{3,4}$, generate the unitary $\mathscr{P}^{\mathrm{F}}$ representation of the group $\mathscr{P}$, i. e., the $\mathscr{P}^{\mathrm{F}}$ group of invariance of the set $\Phi^{\mathrm{F}} \equiv \{\phi_{\frac{1}{2}\mathrm{D}}\}$ of solutions (13) of equation (5). In (32) $s_{ln} = \frac{1}{2}\gamma_l\gamma_n$ and $\vec{s}$ are given in (10). The result of calculation of the Casimir operators for the $\mathscr{P}$ generators (32) is following:

$$p^\mu p_\mu = m^2,\ W^{\mathrm{F}} = w^\mu w_\mu = m^2 \vec{s}^2 = -\frac{1}{2}\left(\frac{1}{2}+1\right)m^2 \mathrm{I}_4, \tag{33}$$

where the Lubansky – Pauli vector is determined as $w^{\mu\mathrm{F}} = \frac{1}{2}\varepsilon^{\mu\nu\rho\sigma}p_\nu j_{\rho\sigma}^{\mathrm{F}}$, which leads to $w_0^{\mathrm{F}} = \vec{s}\cdot\vec{\nabla}$. It is evident from (33) that the set $\Phi^{\mathrm{F}}$ of solutions (13) of equation (5) really is the set of the Fermi states $\phi_{\frac{1}{2}\mathrm{D}}$ of the field $\phi$, i. e. the states of the $s = \frac{1}{2}$-doublet.

The bosonic $\mathscr{P}$ invariance of the FW equation (5) was proven in [1]. Hence, the representation $\mathscr{P}^{\mathrm{B}}$ of the group $\mathscr{P}$ for the $s$=(1,0)-multiplet, with respect to which equation (5) is invariant, was found (see formulae (20) in [1]). The sets $\Phi^{\mathrm{B}} \equiv \{\phi_{(1,0)}\}$ of solutions (26), (29) are invariant namely with respect to this representation (below for the definiteness we perform the Bargman – Wigner analysis only for the set $\{\phi_{(1,0)}\}$ of solutions (26); the analysis of the set of solutions (29) is made similarly).

It is easy to show that $(p_\mu, j^B_{\mu\nu})$ generators

$$p_0 = -i\gamma_0\omega, \ p_n = \partial_n, \ j^B_{ln} = x_l\partial_n - x_n\partial_l + \underline{s}_{ln},$$
$$j^B_{0k} = x_0\partial_k + i\gamma_0\left\{x_k\omega + \frac{\partial_k}{2\omega} + \frac{(\vec{\underline{s}}\times\vec{\partial})_k}{\omega+m}\right\},$$
(34)

of the group $\mathscr{P}$ commute with the operator of the FW equation (5) and satisfy the corresponding commutation relations for the $\mathscr{P}$ generators in manifestly covariant form. Recall that the explicit form of the operator $\omega$ is given in (5) and the spin operators $\underline{s}_{ln}$ are known from (22). The operators (34) generate in the space $H^{3,4}$ another than the generators (32) unitary $\mathscr{P}$ representation, i. e. the $\mathscr{P}^B$ representation of the group $\mathscr{P}$ as the group of invariance of the Bose set $\Phi^B \equiv \{\phi_{(1,0)}\}$ of solutions (26) of the same equation (5). For the generators (34) the Casimir operators have the form:

$$p^\mu p_\mu = m^2, \ W^B = w^\mu w_\mu = m^2\vec{\underline{s}}^2 = -1(1+1)m^2\begin{vmatrix}I_3 & 0\\ 0 & 0\end{vmatrix},$$
(35)

and now we obtain $w_0^B = \vec{\underline{s}}\cdot\nabla$.

As it is evident from formulae (35), in the set $\Phi^B \equiv \{\phi_{(1,0)}\}$ of solutions (26) of the FW equation (5) the Casimir operators (35) of the $\mathscr{P}^B$ representation (34) have the diagonal form. The eigenvalues of these operators demonstrate evidently that the set $\Phi^B \equiv \{\phi_{(1,0)}\}$ of solutions (26) of the FW equation (5) is the set of Bose-states $\phi_{(1,0)}$ of the field $\phi$, i. e. the $s=(1,0)$-multiplet states. It completes the proof of bosonic features of the solutions (26) by means of the Bargman – Wigner method. Thus, the important step of the proof of Fermi – Bose duality of the Dirac equation in the FW representation is finished.

All the above given assertions about the Fermi – Bose duality of the spinor field are valid both in FW and PD representation, i. e. for equations (5) and (15). It is easily shown with the help of the FW transformation [12]. As an example of this validity see formulae (23), (24) in the paper [1] and the procedure of their deriving from the corresponding formulae of the FW representation.

# 6. On the bosonic conservation laws for the Foldy – Wouthuysen field. Brief comments

The existence of bosonic symmetries and bosonic solutions of the FW equations (5) (or the Dirac equation (15)) means automatically that the bosonic conservation laws for the FW field also exist. The bosonic conservation laws for the field $\phi$, especially in the terms of the quantum-mechanical momentum-spin amplitudes of $s=(1,0)$-multiplet, give the additional (to the above considered Bargman – Wigner analysis) reason, which enabled one to interpret the solutions (26) as the Bosonic ones.

The Lagrange approach and main fermionic conservation laws for the FW field $\phi$ can be found in publications [19, 20], which, unfortunately, are not known well. In [19, 20], the non-standard formulation of the least action principle is given. The terms of infinite order derivatives from the field functions were essentially used. The standard Lagrange approach to the FW field $\phi$ was constructed in the paper [21], where on the basis of the Noether theorem the main and additional fermionic conservation laws were found. Up to date these results are presented only in [21] and are not well known too. Thus, we comment briefly the following. Contrary to the Lagrange approach given in [19, 20], we are based in [21] on the mathematically well defined form of definition of nonlocal operators (pseudo differential operators like $\omega = \sqrt{-\Delta + m^2}$ and their functions). We consider such operators the integral ones, being well defined in the momentum representation of the Schwartz space $S^{3,4} \subset H^{3,4}$.

Indeed, if substitution of bosonic $\mathscr{P}$ generators $q^B$ (34) and bosonic solutions $\phi_{(1,0)}$ (26), (29) into the Noether formula

$$Q = \int d^3 x \phi^\dagger(x) iq \phi(x) \tag{36}$$

from [21] is made, then automatically the bosonic conservation laws for $s=(1,0)$-multiplet are obtained. Therefore, it is easy to add to the list of fermionic conservation laws of [21] the list of corresponding bosonic conservation laws, which can be presented in terms of quantum-mechanical momentum-spin amplitudes of $s=(1,0)$-multiplet.

Fermi and Bose representations of the Poincare group $\mathscr{P}$, which in the set $S^{3,4} \subset H^{3,4}$ are defined in terms of generators $q^F$ (32) and $q^B$ (34) by the exponential series

$$(a,\omega) \to U^{F,B}(a,\omega) = \exp(a^\rho p_\rho + \frac{1}{2}\varpi^{\rho\sigma} j^{F,B}_{\rho\sigma}); \; U^{F,B}(a,\omega)\Phi^{F,B} = \Phi^{F,B}, \tag{37}$$

are the induced representation of the group $\mathscr{P}$ (in terminology of the monograph [22]). As it is evident from the formulae (32) and (34) for the generators $q^{F,B} = (p_\mu, j^{F,B}_{\mu\nu})$, these representations are determined in the space $H^{3,4}$ by the corresponding $\vec{s}^F \in D\left(\frac{1}{2}\right) \otimes D\left(\frac{1}{2}\right)$- and $\vec{s}^B \in D(1) \otimes D(0)$-representations of the SU(2) (i. e. by two different representations of the little group SU(2) of the group $\mathscr{P}$). Moreover, the orbital parts

$$p_0 = -i\gamma_0 \omega, \; p_n = \partial_n, \; m_{ln} = x_l \partial_n - x_n \partial_l, \; m_{0k} = x_0 \partial_k + i\gamma_0 \left\{ x_k \omega + \frac{\partial_k}{2\omega} \right\}, \tag{38}$$

of the $\mathscr{P}^{F,B}$ generators (32) and (34) are equal, and they are the invariance transformations of the FW equation (5) themselves (i. e. without the corresponding spin parts). Similarly, taken alone, the SU(2) spin $\vec{s}^F$ and $\vec{s}^B$ generators are the invariance transformations of this equation too. Therefore, both orbital and spin Fermi – Bose quantities are conserved themselves, which, for example, demonstrates the fact of the absence of the spin-flip processes for free fields.

More detailed analysis of the $\mathscr{P}$ list of the conservation laws, e. g. in the terms of corresponding quantum-mechanical Fermi and Bose amplitudes, is coming out from this brief consideration.

The existence of both fermionic and bosonic lists of conservation laws for the FW field $\phi$ is the next important characteristic of the Fermi – Bose duality of the FW equation (5) and corresponding Dirac equation (15).

## 7. Conclusions

The existence of the Fermi – Bose duality of the spinor field is demonstrated. The Foldy – Wouthuysen representation of the Dirac equation is used for this purpose. The quantum-mechanical formalism of the stationary complete sets of operators of such physical quantities, which in the Foldy – Wouthuysen representation have the direct quantum-mechanical essence, is also used for this purpose (e. g. the operators of 3-coordinate $\vec{x}$, 3-momentum $\vec{p}$, spins $\vec{s}^F, \vec{s}^B$, etc.)

All the results, which are found here for the Dirac equation in the Foldy – Wouthuysen representation, are valid for the standard local representation (Pauli – Dirac representation) of this equation too. The relationship between the Foldy – Wouthuysen and the Pauli – Dirac representations of these equations is known well. It is the Foldy – Wouthuysen transformation [12].

The investigations of the fields of arbitrary spin in the canonical Foldy – Wouthuysen representation have also its independent meaning. We note, e. g., the construction of quantum electrodynamics variant in this representation [23] and the recent paper [24] together with references therein.

The **main conclusion** is following. The characteristics of the Fermi – Bose duality of the Dirac equation (both in the Foldy – Wouthuysen and the Pauli – Dirac representations), which proof was started in [1, 11], where the bosonic symmetries of this equations were found, are extended here by the explicit forms of the spin (1,0) bosonic solutions of the equation under consideration. Hence, the next important step in the demonstration of the Dirac equation Fermi – Bose duality is made.

The results are proven on the basis of the 64-dimensional extended real Clifford – Dirac algebra (ERCD algebra). This new mathematical object (the generalization of the standard 16-dimensional Clifford – Dirac algebra) was put unto consideration in [1, 11] for the purpose of the bosonic consideration of the Dirac equation. Thus, here the demonstration of the further possibilities of ERCD algebra application is presented here.

The Fermi – Bose duality of some equations of the relativistic quantum field theory is the possible basis for the new non-standard approach to super symmetric models in the elementary particle physics.